\documentclass[fleqn,usenatbib]{mnras}

\usepackage{newtxtext,newtxmath}

\usepackage[T1]{fontenc}
\usepackage{float}
\usepackage{lscape}
\usepackage{chngpage}
\DeclareRobustCommand{\VAN}[3]{#2}
\let\VANthebibliography\thebibliography
\def\thebibliography{\DeclareRobustCommand{\VAN}[3]{##3}\VANthebibliography}

\usepackage{gensymb}
\usepackage{graphicx}	
\usepackage{amsmath}
\usepackage{physics}
\usepackage{ulem}
\usepackage{scalerel}

\newcommand{\Wmx}{W_\textrm{mx}}
\newcommand{\Wmxc}{W_\textrm{mx}^c}
\newcommand{\parTF}{\theta_{\rm TF}}
\newcommand{\parPV}{\theta_{\rm PV}}
\usepackage{multirow}

\def\la{\mathrel{\hbox{\rlap{\hbox{\lower4pt\hbox{$\sim$}}}\hbox{$<$}}}}
\def\ga{\mathrel{\hbox{\rlap{\hbox{\lower4pt\hbox{$\sim$}}}\hbox{$>$}}}}

\newcommand{\kms}{{\,km\,s$^{-1}$}}
\newcommand{\kmsMpc}{\,km\,s$^{-1}$\,Mpc$^{-1}$}

\newcommand{\HI}{\mbox{\normalsize H\thinspace\footnotesize I}}

\def\apj    {{ApJ }}
\def\apjl   {{ApJL }}
\def\apjs   {{ApJS }}

\title[An improved Tully-Fisher estimate of \texorpdfstring{$H_0$}{}]{An improved Tully-Fisher estimate of \texorpdfstring{$H_0$}{}}

\author[P.\,Boubel et~al.]{
Paula Boubel$^{1}$\thanks{E-mail: paula.boubel@anu.edu.au},
Matthew Colless$^{1}$,
Khaled Said$^{2}$
and Lister Staveley-Smith$^{3}$
\\
$^{1}$Research School of Astronomy and Astrophysics, The Australian National University, Mount Stromlo Observatory, Canberra, ACT 2611, Australia\\
$^{2}$School of Mathematics and Physics, University of Queensland, Brisbane, QLD 4072, Australia\\
$^{3}$International Centre for Radio Astronomy Research (ICRAR), University of Western Australia, 35 Stirling Hwy, Crawley, WA 6009, Australia}

\date{Accepted 2024 August 7. Received 2024 August 7; in original form 2024 July 17}

\pubyear{2024}

\begin{document}
\label{firstpage}
\pagerange{\pageref{firstpage}--\pageref{lastpage}}
\maketitle

\begin{abstract}
We propose an improved comprehensive method for determining the Hubble constant ($H_0$) using the Tully-Fisher relation. By fitting a peculiar velocity model in conjunction with the Tully-Fisher relation, all available data can be used to derive self-consistent Tully-Fisher parameters. In comparison to previous approaches, our method offers several improvements: it can be readily generalised to different forms of the Tully-Fisher relation and its intrinsic scatter; it uses a peculiar velocity model to predict distances more accurately; it can account for all selection effects; it uses the entire dataset to fit the Tully-Fisher relation; and it is fully self-consistent. The Tully-Fisher relation zero-point is calibrated using the subset of galaxies with distances from absolute distance indicators. We demonstrate this method on the Cosmicflows-4 catalogue $i$-band and $W1$-band Tully-Fisher samples and show that the uncertainties from fitting the Tully-Fisher relation amount to only 0.2\kmsMpc. Using all available absolute distance calibrators, we obtain $H_0=73.3$\,$\pm$\,2.1\,(stat)\,$\pm$\,3.5\,(sys)\kmsMpc, where the statistical uncertainty is dominated by the small number of galaxies with absolute distance estimates. The substantial systematic uncertainty reflects inconsistencies between various zero-point calibrations of the Cepheid period--luminosity relation, the tip of the red giant branch standard candle, and the Type Ia supernova standard candle. However, given a reliable set of absolute distance calibrators, our method promises enhanced precision in $H_0$ measurements from large new Tully-Fisher samples such as the WALLABY survey.
\end{abstract}



\section{Introduction}
\label{sec:introduction}

The Tully-Fisher relation has been used to measure the Hubble constant ($H_0$) since its discovery by \citet{Tully_1977}. They found a preliminary result of $H_0$\,=\,80\kmsMpc, the most accurate at the time (though they did not provide an uncertainty). This determination of $H_0$ comes from comparing inferred distances based on the Tully-Fisher relation to measured redshifts. It typically requires the prior calibration of the Tully-Fisher relation using a sample of galaxies with known distances. The most recent estimates of $H_0$ based on this procedure are listed in Table~\ref{tab:H0tf}. 

Cosmicflows-4 \citep[CF4;][]{Kourkchi_2020, Kourkchi_2022} is the largest Tully-Fisher catalogue to date, containing a full-sky sample of $\sim$10,000 galaxies drawn from various datasets with \HI\ line widths, redshifts, and optical or infrared photometry. For galaxies in selected clusters, K20 found $H_0$\,=\,75.9\,$\pm$\,1.3\kmsMpc\ and $H_0$\,=\,76.2\,$\pm$\,0.9\kmsMpc\ from the Tully-Fisher relations in the $i$ band and $W1$ band, respectively, with an estimated systematic error of $\pm$\,2.3\kmsMpc. For the full CF4 catalogue, including both cluster and field galaxies, \citet{Kourkchi_2020} derived $H_0=75.1\pm0.2$\kmsMpc, with an estimated systematic error of $\pm$\,3\kmsMpc. Using the baryonic Tully-Fisher relation, \citet{Kourkchi_2022} found $H_0=75.5\pm2.5$\kmsMpc.

\citet[][hereafter K20]{Kourkchi_2020} calibrated the Tully-Fisher relation in a series of steps, determining one parameter at a time from subsamples of galaxies. The slope was first calibrated using galaxy clusters, followed by the zero-point using nearby galaxies with known distances. The curvature of the relation was then estimated separately. One problem with this formulation is that it is not guaranteed to be fully self-consistent: the parameters are not fit simultaneously, so parameters fitted later in the procedure are conditional on those fitted previously; this requires some awkward iteration of the procedure to obtain approximate convergence and makes it difficult to estimate statistical errors. Another problem is that K20 did not account for the intrinsic scatter in the Tully-Fisher relation in their fitting procedure, leading to under-estimation of the uncertainties for individual data points and potentially biasing the fits. Finally, in comparing their results from different passbands, K20 found systematic differences that called for manual adjustments. 

In this work, we present an improved method for determining $H_0$ from CF4 and other Tully-Fisher datasets. This is derived from the novel methodology developed by \citet{Boubel_2024} for Tully-Fisher distance estimation. This uses Bayesian forward-modelling to obtain distances for individual galaxies simultaneously with the parameters of models for the Tully-Fisher relation and the peculiar velocity field. The method produces fully self-consistent results for all these quantities in a single-step process. The resulting \textit{relative} distances are $h$-dependent (where $h\equiv H_0/100$), so a $H_0$ value is required in order to obtain the \textit{absolute} distances. Conversely, by calibrating the absolute distance scale (i.e.\ the zero-point of the Tully-Fisher relation) using known distances for a subset of galaxies, the method can yield a measurement of $H_0$.

We will show that our improved method reduces many of the statistical and systematic uncertainties in fitting the Tully-Fisher relation. The remaining errors are dominated by problems with the absolute distance indicators used to calibrate the Tully-Fisher relation zero-point. These zero-point errors have two sources: a significant statistical uncertainty that is mainly due to the small number of galaxies with absolute distance estimates, and a substantial systematic zero-point uncertainty that is apparent from the inconsistencies between different absolute calibrators. Our improved method for deriving $H_0$ from the Tully-Fisher relation will realise its full potential once these issues with absolute distance estimators are resolved.

This paper is organised as follows. In Section~\ref{sec:TFR+PVmodel} we summarise the methodology and results of \citet{Boubel_2024} for jointly fitting the Tully-Fisher relation and a peculiar velocity model, and thereby estimating galaxy distances. We describe refinements to this method that we have applied for the analysis here, such as improvements to the sample selection function. In Section~\ref{sec:zpcalibration} we calibrate the zero-point of the Tully-Fisher relation using two different techniques: absolute calibration with galaxies of known distance and relative calibration to the Tully-Fisher relation obtained by K20 for the same sample. In Section~\ref{sec:H0measurement} we describe the method we use to estimate $H_0$ and present our results. In Section~\ref{sec:errors} we examine the various sources of statistical and systematic uncertainty affecting our measurements of $H_0$. We summarise our conclusions in Section~\ref{sec:conclusions}.

\begin{table}
\centering
\caption{Recent $H_0$ estimates using the Tully-Fisher relation.}
\label{tab:H0tf}
\begin{tabular}{lcc}
\hline\hline
Reference & $H_0$ [\kmsMpc\,]\\
\hline
\citet{Russell_2009} & 84.2$\pm$6.0 \\
\citet{Hislop_2011} & 79.0$\pm$2.0 \\
\citet{Sorce_2012} & 75.2$\pm$3.0 \\
\citet{Sorce_2013} & 74.0$\pm$4.0 \\
\citet{Neill_2014} & 74.4$\pm$1.0 \\
\citet{Sorce_2014} & 75.2$\pm$3.3 \\
\citet{Schombert_2020} & 75.1$\pm$2.3 \\
\citet{Kourkchi_2020} & 75.1$\pm$0.2 \\
\citet{Kourkchi_2022} & 75.5$\pm$2.5 \\
\citet{Courtois_2023} & 74.5$\pm$1.0 \\
\hline
\end{tabular}
\end{table}

\section{Combined Tully-Fisher relation and peculiar velocity model fits}
\label{sec:TFR+PVmodel}

\subsection{Magnitude selection functions}
\label{sec:magselfn}

Owing to the use of the forward Tully-Fisher relation and the conditional probability distribution by \citet{Boubel_2024}, selection function factors independent of magnitude cancel out. While the CF4 Tully-Fisher sample is not directly affected by the magnitude limits of the optical/infrared photometry, the \HI\ flux limit indirectly imposes a selection on apparent optical/infrared magnitude \citep{Kourkchi_2020}. This effective magnitude selection function $F(m)$ was determined empirically by comparing the observed distribution of magnitudes in the sample with the expected distribution in the absence of selection effects. \citet{Boubel_2024} modelled it as a linear drop-off in galaxy number counts after some magnitude $m_1$. Here we refine this model because inaccuracies in the magnitude selection function manifest as residual statistical biases. For example, $H_0$ measurements increasing with redshift could hint at the presence of Malmquist bias due to scatter in the distance indicator relation combined with a magnitude or flux limit \citep{Strauss_1995} or could be due to living in a local over-dense region. Our method is not subject to this bias provided both the magnitude selection function and the intrinsic scatter are sufficiently accurately characterised. 

\begin{figure}
\centering
\includegraphics[width=1.0\columnwidth]{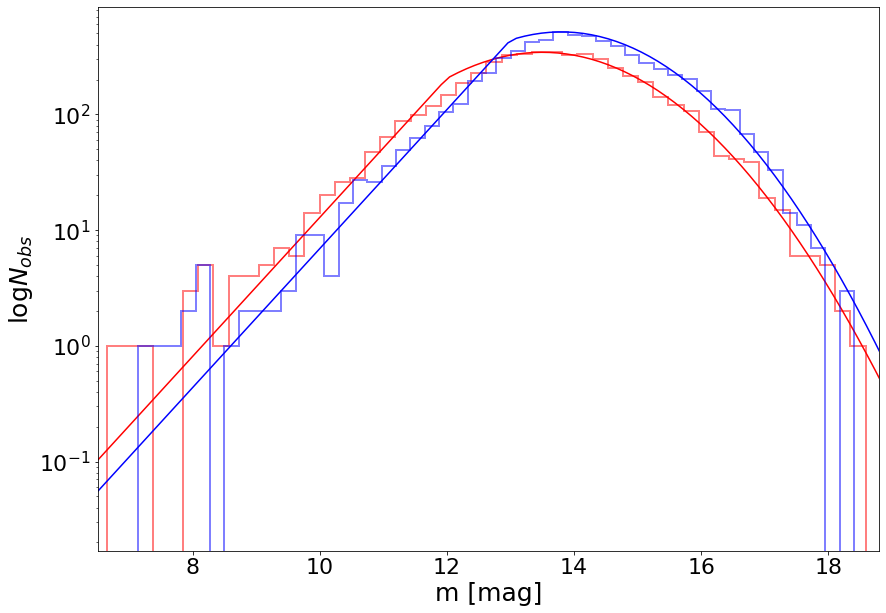}
\caption{Logarithm of observed number counts of galaxies for different apparent magnitudes in the $i$-band (blue) and $W1$-band (red). The adopted models are shown as solid lines in the respective colours.}
\label{fig:Nobs}
\end{figure}

Figure~\ref{fig:Nobs} shows the distributions of magnitudes in the $i$-band and $W1$-band. The bright end (up to $m=m_1$) corresponds to a magnitude-complete sample, where we expect the true number of galaxies to increase with magnitude as $N_{\rm true}(m) = N(m_1)10^{0.6(m-m_1)}$ to a good approximation at these low redshifts. Beyond $m_1$, sample incompleteness causes the observed number of galaxies to diverge from this relation. We model the fall-off in completeness as a Gaussian with a peak at $m_2$, given by $N_\textrm{obs}(m) = N(m_2)10^{a(m-m_2)^2}$. The effective selection function for $m$ can thus be approximated as
\begin{align} 
\log & F(m) = \log N_{\rm obs}(m) - \log N_{\rm true}(m) \nonumber \\
     & =
\begin{cases}
0 & m \leq m_1 \\
a(m-m_2)^2-a(m_1-m_2)^2 - 0.6(m-m_1) & m_1 < m
\end{cases}
\label{eqn:maglimsoft}
\end{align}
where for SDSS we empirically determine $m_1=13.0$, $m_2=13.8$ and $a=-0.11$, while for WISE we find $m_1=12.0$, $m_2=13.5$ and $a=-0.10$. These completeness models are shown in Figure~\ref{fig:Nobs}.

Any residual errors in the selection function would translate into additional systematic errors. However, these are expected to be negligible compared to other statistical and systematic uncertainties, as tests in which $m_1$ and $m_2$ were varied by up to a magnitude produced no observable systematic differences in estimating $H_0$.

\subsection{Fitting methodology}
\label{sec:fitmethod}

The Tully-Fisher relation is a relative distance indicator that must be calibrated using absolute distance indicators from other rungs of the distance ladder. However, this calibration only affects the zero-point of the relation (see Section~\ref{sec:zpcalibration}); the remaining parameters of the Tully-Fisher model can all be determined purely from the Tully-Fisher dataset. One way to do this is by assuming that galaxies belonging to the same cluster all have the same distance. K20 applied this method to the Tully-Fisher CF4 sample to calibrate the Tully-Fisher parameters (excluding the zero-point). 

In \citet{Boubel_2024}, we performed the calibration using a peculiar velocity model to predict galaxy distances. The parameters of the peculiar velocity model are determined simultaneously with the parameters of the Tully-Fisher relation. This method has the following advantages: (i)~it exploits the entire dataset to determine the Tully-Fisher model parameters, rather than only cluster galaxies; (ii)~it is a fully self-consistent, single-step process, rather than requiring iteration to achieve approximate convergence; and (iii)~it is readily adapted to modified forms of the Tully-Fisher relation. 

\begin{table}
\centering
\caption{Parameters of the Tully-Fisher relation and peculiar velocity field fitted to the Cosmicflows-4 SDSS $i$-band and WISE $W1$-band samples.}
\label{tab:params}
\begin{tabular}{lcc}
\hline\hline
Parameter & SDSS $i$-band & WISE $W1$-band\\
\hline
$a_0(h=1)$ & $-$20.403\,$\pm$\,0.006 & $-$19.954\,$\pm$\,0.007 \\
$a_1$ & $-$8.22 $\pm$ 0.05 & $-$9.62 $\pm$ 0.07 \\
$a_2$ & 9.2 $\pm$ 0.3 & 9.8 $\pm$ 0.4 \\
$\epsilon_0$ & 0.622 $\pm$ 0.003 & 0.362 $\pm$ 0.004 \\
$\epsilon_1$ & $-$1.01 $\pm$ 0.02 & $-$0.63 $\pm$ 0.03 \\
$\beta$ & 0.33$\pm$0.03 & 0.36$\pm$0.02 \\
$V_x$ [\kms] &  ~~$-$203$\pm$35 &  $-$225$\pm$15 \\
$V_y$ [\kms] & ~~7$\pm$13 &~~~~$-$7$\pm$13 \\
$V_z$ [\kms] & ~~$-$65$\pm$20 & ~~$-$46$\pm$16 \\
\hline
\multicolumn{3}{l}{$\mathbf{V}_{\textrm{ext}}$ is in Supergalactic Cartesian coordinates and the CMB frame.}	
\end{tabular}
\end{table}

A full description of the method is provided in \citet{Boubel_2024}. Briefly, we estimate model parameters by maximising the likelihood constructed as the product (over all sample galaxies $i$) of the conditional probability of observing an apparent magnitude $m$ given the other observables (position ($\alpha$,$\delta$), redshift $z$, velocity width $w$) and the parameters of the Tully-Fisher model ($\parTF$) and the peculiar velocity model ($\parPV$). This conditional probability is
\begin{equation}
P(m\,|\,w,z,\alpha,\delta,\parTF,\parPV) = \frac{F(m)\exp\left[-\frac{(m-m^\prime)^{2}}{2\sigma_\textrm{TF}^2}\right]}{\int F(m)\exp\left[-\frac{(m-m^\prime)^{2}}{2\sigma_\textrm{TF}^2}\right]\,dm}
\label{eqn:mcondprob}
\end{equation}
where $m^\prime$ is the predicted apparent magnitude
\begin{align}
m^\prime(w,z,\alpha,\delta,\parPV,&\parTF) = \nonumber \\
&M^\prime(w) + 25 + 5\log(1+z) + 5\log d_C(z_c^\prime) ~.
\label{eqn:mpred}
\end{align}
This prediction uses both the Tully-Fisher model for the absolute magnitude given the velocity width, $M^\prime(w)$, and the peculiar velocity model for the cosmological redshift, $z_c$, given by
\begin{equation}
z_c^\prime(z,\alpha,\delta,\parPV) = \frac{z-z_p^\prime}{1+z_p^\prime} 
 = \frac{z-u(z,\alpha,\delta,\parPV)/c}{1+u(z,\alpha,\delta,\parPV)/c} ~.
\label{eqn:zcpred}
\end{equation}
The peculiar velocity model $u(z,\alpha,\delta,\parPV)$ is parameterised as
\begin{equation}
u(z,\alpha,\delta,\parPV) = \beta V_{\textrm{pred}}(z,\alpha,\delta) + \mathbf{V}_\textrm{ext} \cdot {\mathbf{\hat{r}}(z,\alpha,\delta)} 
\label{eqn:velocitymodel}
\end{equation}
where $\beta$ is the velocity scaling and $\mathbf{V}_\textrm{ext}$ is the external dipole \citep[see][]{Boubel_2024}.

The selection function's dependence on magnitude is characterised by $F(m)$ and accounted for in Equation~\ref{eqn:mcondprob}; other factors in the selection function cancel between the numerator and denominator. This is one of the advantages of using the conditional probability, which also avoids the need to characterise the intrinsic velocity width distribution and the redshift distribution as a function of sky position.

The log-likelihood is constructed from Equation~\ref{eqn:mcondprob} as
\begin{align}
\ln \mathcal{L} &= \sum_i \ln P(m\,|\,w,s,z,\alpha,\delta,\parTF,\parPV) \nonumber \\
 &= \sum_i -\ln\left(\sqrt{2\pi}\sigma_{\rm TF}\right)-\left[\frac{(m-m^\prime)^2}{2\sigma_{\rm TF}^2}\right] ~.
\label{eqn:lnlike}
\end{align}
We can then use a Markov chain Monte Carlo (MCMC) algorithm \citep{emcee} to sample this likelihood and simultaneously derive the best-fit parameters for the Tully-Fisher model ($\parTF$) and the peculiar velocity model ($\parPV$). 

\begin{figure}
\centering
\includegraphics[width=1.0\columnwidth]{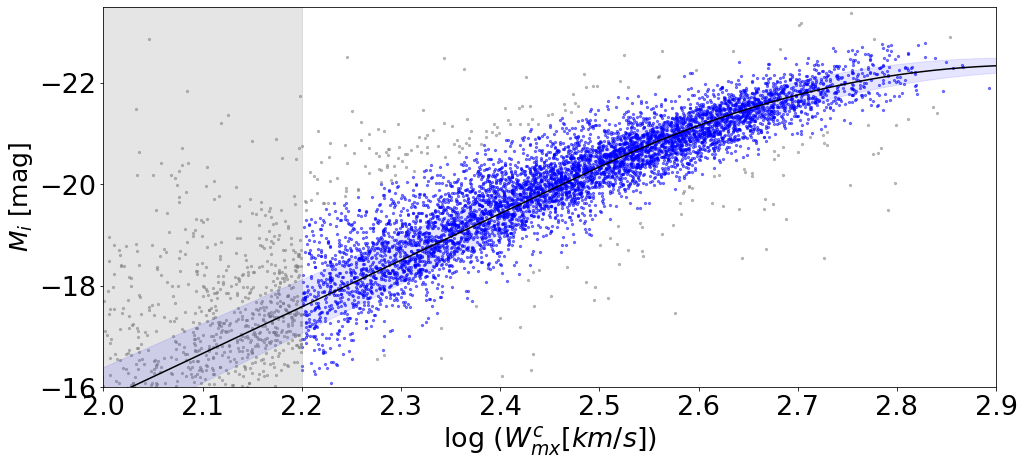}
\includegraphics[width=1.0\columnwidth]{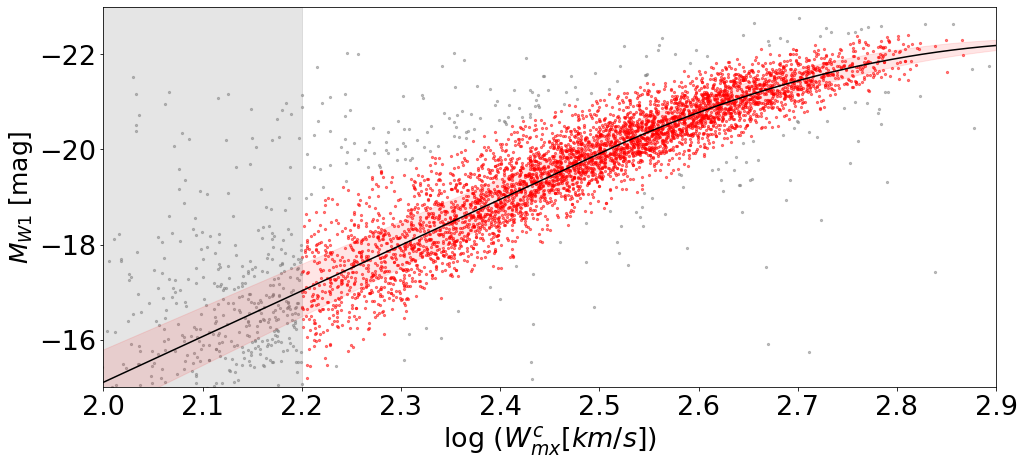}
\caption{Best-fit $i$-band (top/blue) and $W1$-band (bottom/red) CF4 Tully-Fisher relations for absolute magnitudes assuming $H_0$\,=\,100\kmsMpc. Table~\ref{tab:params} gives the fit parameters and uncertainties. The full CF4 datasets are used, excluding galaxies (grey) that have low velocity widths ($\log\Wmxc < 2.2$) or are outliers from the Tully-Fisher relation (residual $>3\sigma_{\rm{TF}}$). Blue/red shaded regions show the intrinsic scatter model and cover $\pm$1$\sigma_{\rm{TF}}$.}
\label{fig:tfr_full}
\end{figure}

\subsection{Best-fitting models}
\label{sec:bestfitmodels}

We apply the selection function and fitting methodology above to fit the CF4 datasets in the $i$-band and the $W1$-band. We exclude from these fits galaxies that have low velocity widths ($\log\Wmxc < 2.2$; 1163 galaxies in $i$ and 563 galaxies in $W1$) or that are outliers from the Tully-Fisher relation (residual $>3\sigma_{\rm{TF}}$; a further 230 galaxies in $i$ and 240 galaxies in $W1$). This leaves 6084 galaxies in the $i$-band fit and 4639 galaxies in the $W1$-band fit. For the purpose of these fits we computed absolute magnitudes assuming $H_0$\,=\,100\kmsMpc, so the zero-points correspond to $h=1$. The resulting best-fit parameters (and their 1$\sigma$ uncertainties) are listed in Table~\ref{tab:params}; the corresponding best-fit Tully-Fisher models are shown in Figure~\ref{fig:tfr_full}.

\subsection{Galaxy distances}
\label{sec:galaxydistances}

As shown by \cite{Boubel_2024}, the conditional probability of a galaxy having cosmological redshift $z_c$ given the observables and specific model parameters can be written as
\begin{align}
P(z_c\,&|\,m, w,z, \alpha,\delta,\parTF,\parPV) \nonumber \\
 &\propto \frac{F(m)\exp\left[-\frac{(m-m^\prime)^{2}}{2\sigma_\textrm{TF}^2}\right]\exp \left[ -\frac{(cz_c - cz_c^\prime)^{2}}{2\sigma^{2}_{v}} \right]}{\sqrt{2\pi}\sigma_{v}\int F(m)\exp\left[-\frac{(m-m^\prime)^{2}}{2\sigma_\textrm{TF}^2}\right]\,dm} ~.
\label{eqn:zccondprob}
\end{align}
This estimate of $z_c$, and the corresponding distance, combines information from the galaxy's Tully-Fisher relation residual and the prediction of the fitted peculiar velocity model, but is determined primarily by the latter (as $\sigma_v \ll \sigma_\textrm{TF}$).

As noted above, a good consistency test for these distance measurements is to check that $H_0$ remains constant for galaxies in different redshift ranges. We perform this test in Section~\ref{sec:errors}, finding that the inferred distance moduli of individual galaxies show no variation with redshift when compared to a model using the best-fit value $H_0$ (as expected over this redshift range). 

We can also estimate a `Tully-Fisher distance' by just using each galaxy's offset from the best-fit Tully-Fisher relation, with an uncertainty from $\sigma_\textrm{TF}$ (typically $\sim$25\%). In Section~\ref{sec:H0measurement} we take the combined distance estimates as $z_c$ and the Tully-Fisher distance estimates as $d_L$ for each galaxy, and use these to fit $H_0$ for the entire sample.

\section{Calibrating the Tully-Fisher zero-point}
\label{sec:zpcalibration}

The absolute magnitude $M$ of a galaxy can be determined from its apparent magnitude $m$ and its cosmological redshift $z_c$ as
\begin{equation} 
M = m - 5 \log d_L(z_c) - 25
\label{eqn:abs_mag}
\end{equation}
where $d_L$, the luminosity distance, is measured in Mpc. In order to compute a distance from a redshift $z_c$, a cosmological model is required. We assume a flat $\Lambda$CDM cosmology with $\Omega_m$\,=\,0.315, as favoured by the Planck collaboration \citep{2020_Planck} and write $H_0$\,=\,100$h$\kmsMpc. Because the product $H_0 d_L$ is fixed for a given redshift, we can include the $h$-dependence explicitly as
\begin{equation} 
M = m - 5 \log d_L(z_c^\prime, h=1) + 5 \log h - 25 ~.
\label{eqn:abs_mag_h}
\end{equation}
The assumed value of $h$ shifts the predicted absolute magnitude $M$ by a fixed amount for all galaxies in the sample, so changes to $\log h$ offset the intercept of the Tully-Fisher relation and have no effect on the other parameters. This means it is not possible to measure $H_0$ directly from the Tully-Fisher relation unless the true zero-point of the relation is determined by other means.

In this section we describe two methods for setting the zero-point of the Tully-Fisher relation: an absolute calibration using a subset of CF4 Tully-Fisher galaxies with known distances and a calibration relative to that obtained previously for the CF4 Tully-Fisher dataset by K20. In this analysis, the zero-point is fit separately, holding the other Tully-Fisher relation parameters fixed, in order to more easily test the systematics affecting it. As mentioned above, this is possible because the zero-point is independent of the other parameters; in principle, however, the likelihoods can be combined to fit all parameters simultaneously.

\subsection{Absolute calibration}
\label{sec:abscalib}

To calibrate the zero-point of the Tully-Fisher relation, we can take advantage of three independent methods of determining accurate distances: the Cepheid period--luminosity relation (CPLR), the tip of the red giant branch (TRGB) standard candle, and the Type Ia supernova (SNIa) standard candle. These distance measures are only available for small subsets of galaxies within the full CF4 Tully-Fisher sample. In the $i$-band sample, there are 16 galaxies with CPLR distances, 26 galaxies with TRGB distances, and 61 galaxies with SNIa distances (a total of 98 unique galaxies); in the $W1$-band sample, there are 24 galaxies with CPLR distances, 55 galaxies with TRGB distances, and 76 galaxies with SNIa distances (a total of 136 unique galaxies). In the small number of cases where distances are available from multiple sources, the average is taken. We use the homogenised distance moduli from \cite{Makarov_2014}, who set their standard distance scale to be consistent with \citet{Rizzi_2007}, the same as that used by K20, and where the SNIa peak absolute magnitudes were calibrated using Cepheids. Different choices for the CPLR/TRGB/SNIa zero-point scales have a significant effect on the measured value of $H_0$; this will be accounted for as a systematic uncertainty (we discuss this issue in more in detail in Section~\ref{sec:errors}). For this subset of CF4 galaxies, we use the apparent magnitudes $m_i$ or $m_{W1}$ from \citet{Kourkchi_2020} and the measured CPLR/TRGB/SNIa distances to compute absolute magnitudes using Equation~\ref{eqn:abs_mag}. This gives a Tully-Fisher relation in which the distance scale is absolute. 

In \citet{Boubel_2024}, the adopted model for the CF4 Tully-Fisher relation had the form
\begin{equation} 
M =
\begin{cases}
a_0 + a_1w & (w < 0) \\
a_0 + a_1w + a_2w^2 & (w \geq 0) \\
\end{cases}
\label{eqn:TFreln}
\end{equation}
where $w \equiv \log{\Wmxc} - 2.5$ and $w = 0$ is the break-point above which the relation shows curvature. We also adopted a linear model for the scatter in magnitude about the Tully-Fisher relation as a function of velocity width
\begin{equation}
\sigma_\textrm{TF} = \epsilon_{0} + \epsilon_{1} (w + 2.5) 
\label{eqn:TFscatter}
\end{equation} 
where $\sigma_\textrm{TF}$ accounts for both scatter intrinsic to the Tully-Fisher relation and scatter due to unmodelled peculiar velocities (errors in the linear velocity field model or non-linear peculiar motions).

A $H_0$ value of 100\kmsMpc\ was assumed in \citet{Boubel_2024}, so their value for $a_0$ is only correct if $h=1$. We fix the Tully-Fisher model parameters $a_1$, $a_2$, $\epsilon_0$, and $\epsilon_1$ (collectively, $\parTF$) to the values obtained using the method described in Section~\ref{sec:fitmethod} and summarised in Table~\ref{tab:params}. We then fit for the zero-point $a_0$ only, maximising the (logarithm of) the likelihood given by the product (over all calibrator galaxies $i$) of the probability distribution
\begin{equation}
P(m\,|\,w,d_L,\parTF) = \frac{1}{\sqrt{2\pi}\sigma(w)} \exp\left[-\frac{(m-m^\prime)^{2}}{2\sigma(w)^2}\right]
\label{eqn:TFmodel}
\end{equation}
where
\begin{equation}
m^\prime = 5\log d_L + M^\prime(w) + 25
\label{eqn:mprime}
\end{equation}
and $M^\prime(w)$ is given by Equation~\ref{eqn:TFreln}. The scatter in $m-m^\prime$ is denoted $\sigma(w)$ and is the quadrature sum of the intrinsic Tully-Fisher scatter $\sigma_{\rm{TF}}$, the observational errors in absolute magnitudes arising from the velocity width errors $e_w$, and the observational errors in the absolute magnitudes themselves $e_M$ (which are the errors in the apparent magnitudes if assuming a value of $H_0$ and the errors in the distance moduli if using distances from absolute distance estimators). The scatter in $m-m^\prime$ is thus given by
\begin{equation}
\sigma^{2}(w) = \sigma_{\rm{TF}}^{2}(w) + \left( \frac{\dd{M(w)}}{\dd{w}} e_w \right) ^{2} + e_{M}^{2} ~.
\label{eqn:fullscatter}
\end{equation}

\begin{figure}
\centering
\includegraphics[width=1.0\columnwidth]{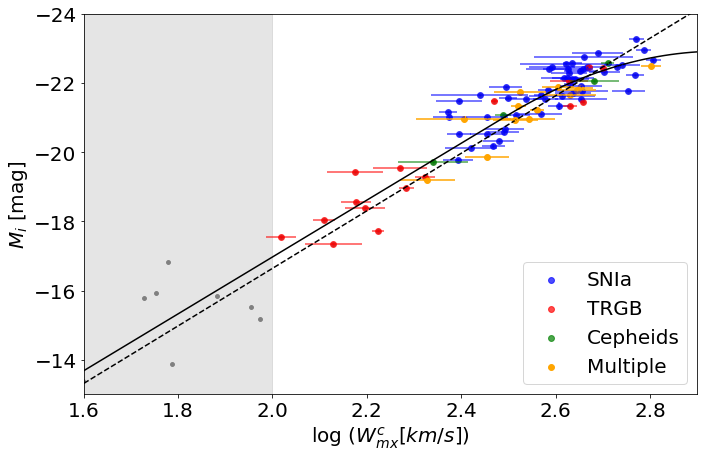}
\includegraphics[width=1.0\columnwidth]{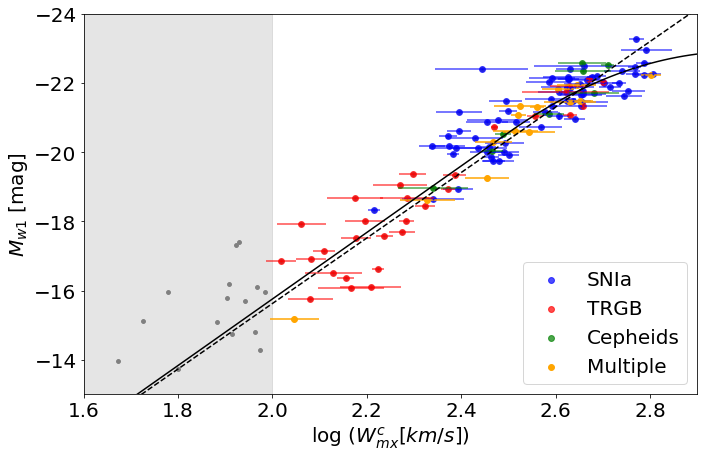}
\caption{Fits (solid curves) to the calibrator galaxies establish an absolute zero-point for the Tully-Fisher relation. Distances come from CPLR (green), TRGB (red) or SNIa (blue), or the average of all available distances (orange). Absolute magnitudes come from the known distances and measured magnitudes in the $i$-band (top panel) and $W1$-band (bottom panel). Errorbars show measurement errors in $\log\Wmxc$; magnitude errors are incorporated in the fitting procedure through the scatter model. K20's linear fits to the same calibrators are shown as dashed lines. In our analysis, galaxies with $\log\Wmxc < 2$ are excluded (grey points); in K20, galaxies with $M>-17$ in the $i$-band and $M>-16.1$ in the $W1$-band are excluded.}
\label{fig:zp_calib}
\end{figure}

This procedure gives a zero-point of $a_0$\,=\,$-$21.09\,$\pm$\,0.06 for the $i$-band and $a_0$\,=\,$-$20.59\,$\pm$\,0.03 for the $W1$-band. Figure~\ref{fig:zp_calib} shows the best-fit Tully-Fisher relations for the calibrator galaxies. K20 obtained $a_0=-$20.80\,$\pm$\,0.1 for the $i$-band and $a_0=-$20.36\,$\pm$\,0.07 for the $W1$-band, but these values cannot be compared directly to our zero-points as the slopes of the Tully-Fisher relations are slightly different. Figure~\ref{fig:zp_calib} shows a comparison of the Tully-Fisher relations used in fitting the calibrator galaxies. 

K20's fitting procedure differs from ours in a few key ways. Firstly, they fit the Tully-Fisher parameters (including the zero-point) using a linear inverse Tully-Fisher relation that does not account for the intrinsic scatter, whereas we have included curvature and a scatter model. One effect of using the curved relation to fit the zero-points is that it allows the linear part of the relation to shift to brighter magnitudes because it accommodates the flattening at high velocity widths, resulting in a more negative $a_0$ (i.e.\ a brighter zero-point magnitude). Secondly, we include SNIa in our absolute distance calibrators for $a_0$, which also gives a brighter zero-point (see Table~\ref{tab:zpvals}). Lastly, in K20 the Tully-Fisher slope was calculated by assuming that galaxies within clusters were at the same distance, computing individual slopes for each galaxy cluster, and iteratively shifting the Tully-Fisher relations until they converged to a universal Tully-Fisher relation. In our method, the Tully-Fisher parameters are all simultaneously fit using the full CF4 sample; distances are not assumed, but predicted by a peculiar velocity model that is fit simultaneously with the Tully-Fisher relation. This approach is conceptually simpler and allows both the inclusion of many more galaxies in the fit and generalisations of the form of the Tully-Fisher relation. 

\subsection{Relative calibration}
\label{sec:relativeH0}

As well as performing an absolute calibration of the Tully-Fisher relation, we can also calibrate our relation in reference to that of K20. Since, for each galaxy in common, we and K20 are aiming to predict the same apparent magnitude from the same velocity width given the same cosmological redshift, we should expect
\begin{align}
M_1(w_i) &+ 25 + 5\log(1+z_i) + 5\log d_{1}(z_{ci}) \nonumber \\
&= M_2(w_i) + 25 + 5\log(1+z_i) + 5\log d_{2}(z_{ci})
\end{align}
where $w_i$, $z_i$ and $z_{ci}$ are the velocity width, observed redshift and cosmological redshift for galaxy $i$, $M_1(w)$ and $M_2(w)$ are the two Tully-Fisher relations, and $d_{1}(z_c) = cf(z_c)/H_1$ and $d_{2}(z_c) = cf(z_c)/H_2$ are the two comoving distance relations (differing only in their effective Hubble constants). Removing all terms in common gives
\begin{equation}
\log(H_1/H_2) = (M_1(w_i) - M_2(w_i))/5
\end{equation}
which is to say that the log-ratio of the two effective Hubble constants is one-fifth of the difference in the predicted absolute magnitudes from the two Tully-Fisher relations. 

Since we have to allow for the scatter in the Tully-Fisher relations, we can obtain the best overall estimate for the log-ratio of the Hubble constants by minimising the $\chi^2$ statistic given by
\begin{equation}
\chi^2 = \sum_i \left[ \frac{[(M_1(w_i) - M_2(w_i))/5 - \log(H_1/H_2)]^2}{\sigma_1^2(w_i) + \sigma_2^2(w_i)} \right]
\end{equation}
where $\sigma_1(w)$ and $\sigma_2(w)$ are given by Equation~\ref{eqn:fullscatter}.

This leads to a variance-weighted estimate of the log-ratio of the Hubble constants given by
\begin{equation}
\langle \log(H_1/H_2) \rangle = \frac{\sigma_M^2}{5} \sum_i \left[ \frac{M_1(w_i) - M_2(w_i)}{\sigma_1^2(w_i) + \sigma_2^2(w_i)} \right]
\end{equation}
where $\sigma_M$ is the standard error in the mean of the absolute magnitude differences given by 
\begin{equation}
\sigma_M^2 = \frac{1}{\sum_i [\sigma_1^2(w_i) + \sigma_2^2(w_i)]^{-1}} ~.
\end{equation}
The standard error in this estimate of $\log(H_1/H_2)$ is $\sigma_M/\sqrt{5}$.

Thus, by comparing our Tully-Fisher relation to that of K20, we can obtain an estimate of $H_0$ relative to their estimate of $H_0 = 75.9 \pm 1.3 {\rm (stat)} \pm 2.3 {\rm (sys)}$ in the $i$ band and $H_0 = 76.2 \pm 0.9 {\rm (stat)} \pm 2.3 {\rm (sys)}$ in the $W1$ band, which they derive from their cluster calibrators. Since we are using galaxies in common, the relative difference in $H_0$ reflects systematic uncertainties in determining the Tully-Fisher relation between our method and that of K20. To minimise the effect of different forms for the Tully-Fisher relation, we only use galaxies with $2.2 < \log\Wmxc < 2.5$, where the Tully-Fisher relation is linear and the offset is nearly constant (3484 $i$-band galaxies and 2281 $W1$-band galaxies). We find the variance-weighted mean of the difference in predicted absolute magnitudes between K20 and ourselves to be 0.08\,mag ($i$ band) and 0.12\,mag ($W1$ band), corresponding to $\langle \log(H_1/H_2) \rangle = 0.016$ and 0.024, so $H_1/H_2$=1.04 and 1.06. Thus K20's Tully-Fisher relation corresponds to a $H_0$ that is 4--6\% higher than ours. The standard error in $\log(H_1/H_2)$ is about 0.0084 and 0.027, corresponding to a nominal uncertainty in $H_1/H_2$ of 1.9\% and 6.2\%. Adding this in quadrature to the statistical uncertainty in $H_0$ from \citet{Kourkchi_2020}, our estimated Hubble constants are $H_0 = 73.2 \pm 1.9{\rm (stat)} \pm 2.3 {\rm (sys)}$\kmsMpc\ in the $i$ band and $H_0 = 72.0 \pm 4.6{\rm (stat)} \pm 2.3 {\rm (sys)}$\kmsMpc\ in the $W1$ band (in each case adopting K20's estimate of the systematic uncertainty). 

The \citet{Boubel_2024} Tully-Fisher relation has several improvements over that of K20, such as incorporating variation in intrinsic scatter and curvature directly into the model. As seen here, our Tully-Fisher model and method for fitting the parameters result in a slightly lower $H_0$ value than that of K20, based solely on the small differences between the fitted Tully-Fisher relations. 

Adopting the above relative estimates for $H_0$, we calculate the Tully-Fisher zero-points $a_0$ as
\begin{equation}
a_0(h) = 5\log{h_{\rm{rel}}} + a_0(h\!\!=\!\!1),
\end{equation}
where $a_0$($h$=1) is the zero-point from fitting the CF4 Tully-Fisher sample assuming $h$=1 (see Table~\ref{tab:params}) and $h_{\rm{rel}}$ is $h$=0.732 for the $i$ band and $h$=0.720 for the $W1$ band (from the relative calibration procedure described above). This gives $a_0$($h$=0.732) $=-21.0\pm0.1$ for the $i$ band and $a_0$($h$=0.720) $=-20.6\pm0.2$ for the $W1$ band. 

\section{\texorpdfstring{$\mathbf{H_0}$}{} measurement}
\label{sec:H0measurement}

\begin{table}
\centering
\caption{$H_0$ estimates in\kmsMpc, with statistical and systematic errors as described in Section~\ref{sec:errors}.}
	\label{tab:H0vals}
	\begin{tabular}{lcc}
        \hline\hline
        Method & SDSS $i$-band & WISE $W1$-band\\
		\hline
		Absolute & 73.3 $\pm$ 2.1 (stat) $\pm$ 3.5 (sys) & 74.5 $\pm$ 1.2 (stat) $\pm$ 2.6 (sys)\\
  		Relative & 74.3 $\pm$ 2.2 (stat) $\pm$ 2.3 (sys) & 73.2 $\pm$ 4.8 (stat) $\pm$ 2.3 (sys)\\
		K20 & 75.9 $\pm$ 1.3 (stat) $\pm$ 2.3 (sys) & 76.2 $\pm$ 0.9 (stat) $\pm$ 2.3 (sys) \\
  \hline
    \end{tabular}
\end{table}

The relation between redshift $z$ and luminosity distance $d_{L}$ is
\begin{equation}
H_0 d_L(z) = c z f(z)
\label{eqn:DLzmod}
\end{equation}
where, for a flat $\Lambda$CDM cosmology, \citet{Chiba_1998} and \citet{Visser_2004} show that $f(z)$ well approximated by
\begin{equation}
f(z) = 1 + z(1-q_0)/2 - z^2(2-q_0-3q_0^2)/6
\label{eqn:zmod}
\end{equation}
where $q_0 = 3\Omega_m/2 -1$ is the deceleration parameter; if $\Omega_m$\,=\,0.315 then $q_0$\,=\,$-$0.5275. 

Luminosity distances can be predicted via the Tully-Fisher relation from apparent magnitudes $m$ and velocity widths $w$ as
\begin{align}
5\log d_L= m - M^\prime(w) - 25 ~.
\label{eqn:dcpred}
\end{align}

Combining Equations~\ref{eqn:DLzmod} and~\ref{eqn:dcpred} gives an expression for $H_0$ that depends only on the observables $m$, $w$, and $z$, the Tully-Fisher model $\parTF$, and the peculiar velocity model $\parPV$, namely
\begin{equation}
\log H_0 = \log(cf(z)z_c^\prime(z,\alpha,\delta,\parPV)) - \frac{1}{5}(m - M^\prime(w,\parTF) - 25)
\label{eqn:logDLz}
\end{equation}
where $z_c^\prime$ is the cosmological redshift predicted by Equation~\ref{eqn:zccondprob} (driven primarily by the peculiar velocity model) and $M^\prime(w)$ is the absolute magnitude predicted by our Tully-Fisher model.

In the previous section we explored two methods for establishing an absolute calibration of the Tully-Fisher relation: using a set of calibrator galaxies with known distances to find the zero-point (Section~\ref{sec:abscalib}) or obtaining a calibration relative to K20 (Section~\ref{sec:relativeH0}). Switching between methods shifts $H_0$ by the same amount for all galaxies. In subsequent figures and discussion the absolute calibration using all available CPLR, TRGB and SNIa distances is adopted; however, Table~\ref{tab:H0vals} compares results from both methods and from K20.

\begin{figure}
\centering
\includegraphics[width=1.0\columnwidth]{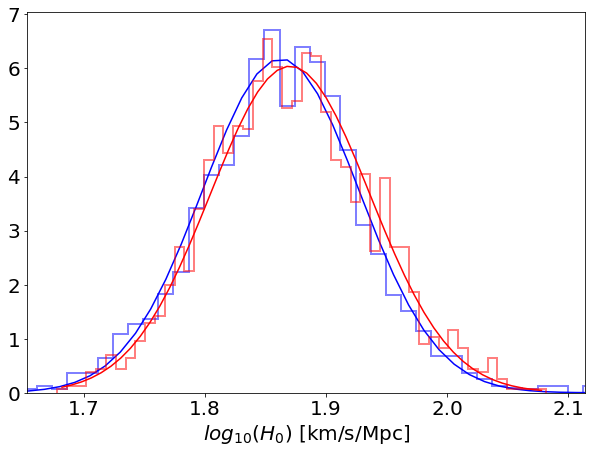}
\includegraphics[width=1.0\columnwidth]{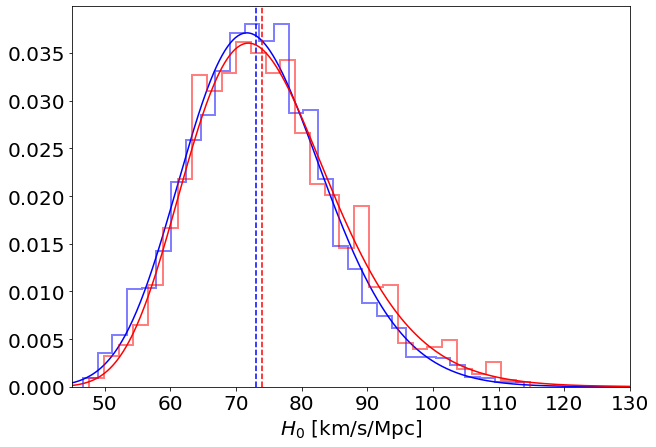}
\caption{The distributions of $\log H_0$ and $H_0$ derived for each galaxy in the CF4 Tully-Fisher sample, excluding those with $\log\Wmxc<2.0$ or $cz<3000$\kms\ (blue for $i$-band, red for $W1$-band). The median of $H_0$ (vertical dashed lines) is 73.0\kmsMpc\ for the $i$-band and 74.0\kmsMpc\ for the $W1$-band; the mode of $H_0$ is 71.7\kmsMpc\ for the $i$-band and 71.9\kmsMpc\ for the $W1$-band.}
\label{fig:H0_dist}
\end{figure}

The uncertainty in $\log H_0$ is predominantly due to the large Gaussian scatter around the Tully-Fisher relation; see Fig.5 of \citet{Boubel_2024}. We thus expect the distribution of $H_0$ to take on a log-normal form. Figure~\ref{fig:H0_dist} shows the distribution of $H_0$ computed using Equation~\ref{eqn:logDLz} for each individual galaxy in the CF4 Tully-Fisher sample. We exclude galaxies with $\log\Wmxc<2.0$ (as for our Tully-Fisher relation fits) or $cz<3000$\kms\ (where peculiar velocities and bulk motions become a large fraction of the Hubble expansion rate, and can introduce large scatter and potential systematics). We see that, except for a few outliers, $\log H_0$ is well-fit by a Gaussian and $H_0$ is well-fit by a log-normal distribution. We perform a 3$\sigma$ clip on $\log H_0$ to remove these outliers, which account for 3.8\% and 3.6\% of the total sample in the $i$-band and $W1$-band respectively. This leaves 5354 galaxies in the $i$-band and 3430 galaxies in the $W1$-band.

\begin{figure}
\centering
\includegraphics[width=1.0\columnwidth]{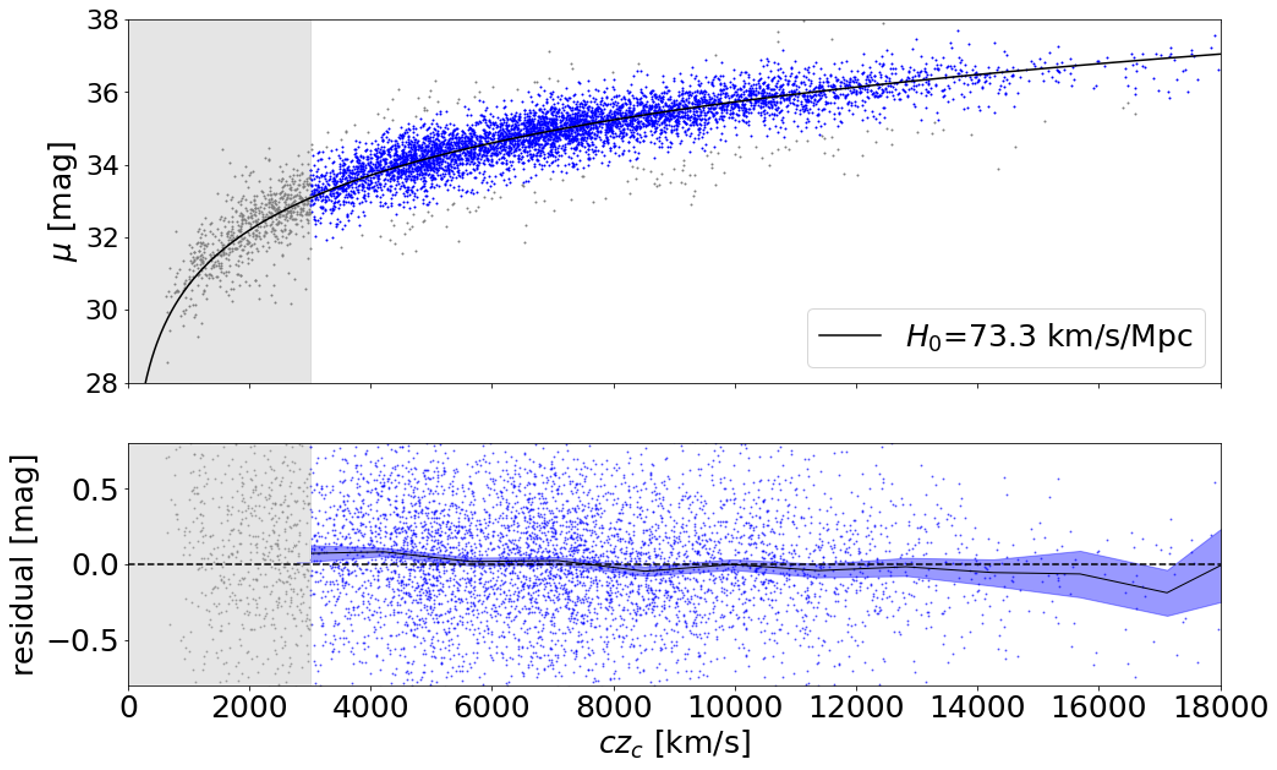}
\includegraphics[width=1.0\columnwidth]{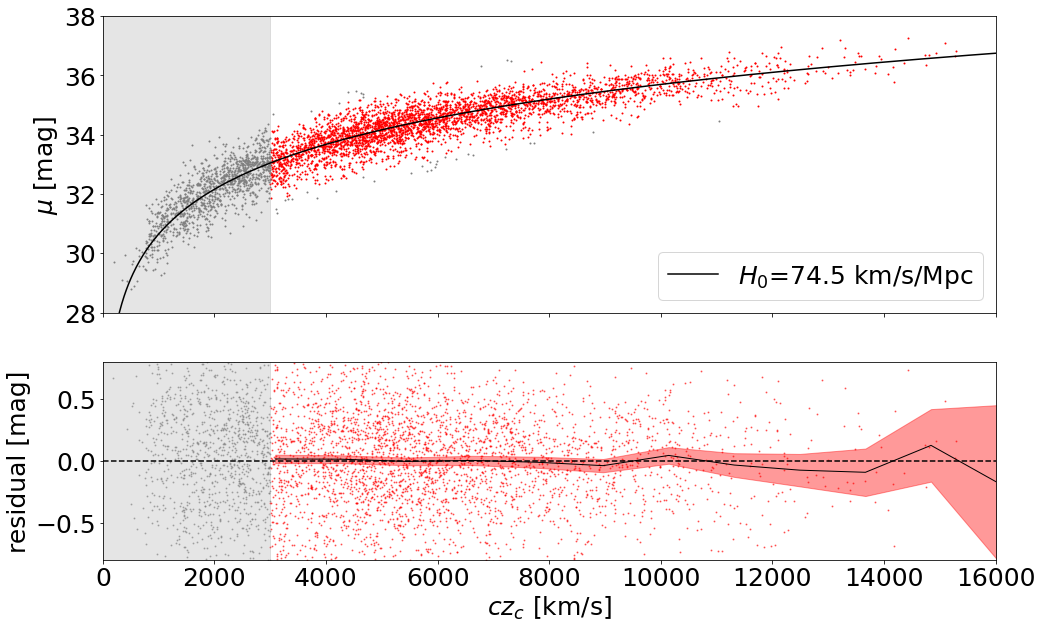}
\caption{Distance moduli of individual galaxies inferred from the Tully-Fisher relation plotted against cosmological redshifts predicted by the peculiar velocity model. The best-fitting values of $H_0$ (black curves) are 73.3$\pm$0.2 for the $i$-band galaxy sample (blue points, upper pair of panels) and 74.5$\pm$0.3 for the $W1$-band galaxy sample (red points, lower pair of panels). Galaxies excluded due to the redshift cut at $cz=3000$\kms\ or the 3$\sigma$ clipping applied to the $\log H_0$ distribution are shown as grey points. After these exclusions, there are 5,354 galaxies in the $i$ band sample and 3,430 galaxies in the $W1$ band sample. The lower panel of each pair shows the residuals of the distance moduli compared to the best-fitting curves, with the median value in redshift bins (solid line) and its associated 1$\sigma$ uncertainty (coloured band).}
\label{fig:hubble_diagram}
\end{figure}

Next, we perform a fit for $H_0$ using Equation~\ref{eqn:fullscatter} for the uncertainty in $M^\prime(w,\parTF)$; the uncertainty in $z_c^\prime$ is negligible in comparison (about 250\kms). Figure~\ref{fig:hubble_diagram} shows the Hubble diagram in terms of the predicted distance modulus, $\mu=25+5\log d_L = m - M^\prime(w)$, against the predicted cosmological redshift $z_c$ for each galaxy; outliers not included in the fit are shown as grey points. The cosmological redshift is used in place of the observed redshift because it is a more precise indicator of true distance, since the peculiar velocities have been corrected using a model. The mean $\log H_0$ value corresponds to the median of the log-normal distribution of $H_0$. These median values are $H_0=73.3\pm0.2$\kmsMpc\ for the $i$-band and $H_0=74.5\pm0.3$\kmsMpc\ for the $W1$-band. These estimates are lower than those obtained by K20, for which there are two primary reasons: first, we use a curved Tully-Fisher relation to fit the galaxies with absolute distance estimates, and second, we include SNIa distances as absolute distance calibrators. Both choices have the effect of decreasing the Tully-Fisher zero-point. Figure~\ref{fig:hubble_diagram} also shows the residuals of the distance moduli compared to the best-fitting curves, together with the median value of the residual in redshift bins and its associated 1$\sigma$ uncertainty; the median residuals reveal no bias and are consistent with zero to within 1.5$\sigma$ over the redshift range. 

The relative calibration gives $H_0=73.2\pm1.4$\kmsMpc\ for the $i$-band and  $H_0=72.0\pm1.0$\kmsMpc\ for the $W1$-band, with an additional systematic uncertainty of $\pm2.3$\kmsMpc\ flowing on from K20's $H_0$ measurements. These estimates are also lower than those of K20. This is a direct result of differences in our Tully-Fisher parameters, which arise either from the data or the model. Differences in the data come down to the galaxy sample, as our method uses about 10 times as many galaxies to calibrate the relation. As for the model, there are several differences. Since our model fits for curvature and slope simultaneously, it allows the linear portion to shift to brighter magnitudes, resulting in a lower $H_0$ value. Additionally, K20 used only measurement errors in the galaxy velocity widths when fitting the inverse Tully-Fisher relation, which tends to over-weight galaxies at higher velocity widths where these errors are smaller. In our method, the model accounts not only for measurement errors in the velocity widths but also for the intrinsic scatter (varying with velocity width) that is known to exist in the Tully-Fisher relation; this also shifts the relation to brighter magnitudes.

\section{Sources of uncertainty}
\label{sec:errors}

The zero-point errors given in Table~\ref{tab:params} indicate the statistical uncertainties in $H_0$ contributed just by fitting the Tully-Fisher relation to the CF4 dataset using the method described in Section~\ref{sec:fitmethod}; they amount to only 0.2\kmsMpc. This section discusses the additional, and substantially larger, sources of statistical and systematic uncertainty that affect the final estimates of $H_0$ given in Table~\ref{tab:H0vals}.

The dominant source of uncertainty in the measurement of $H_0$ arises from the choice of Tully-Fisher zero-point calibrators. For the TRGB distance moduli, especially, there is some contention in the literature regarding the appropriate LMC-based zero-point \citep{Freedman_2020, Hoyt_2021, Yuan_2019}. K20 preferred to use a TRGB calibration consistent with \citet{Rizzi_2007}, instead of using the more recent \citet{Freedman_2019} calibration, because it resulted in Tully-Fisher zero-points (and therefore $H_0$ values) that were consistent with those of the CPLR distances. 

Here, we repeat the absolute zero-point calibration of Section~\ref{sec:abscalib}, but we try different sets of TRGB distance scales based on recent results from the literature. In these fits, all parameters except the Tully-Fisher zero-point are fixed, in order to show how much the zero-point varies from one choice of distance scale to another. A 0.05\,mag difference in the zero-point corresponds approximately to a 1.7\kmsMpc\ (2.3\%) difference in $H_0$. 

Table~\ref{tab:zpvals} shows how the Tully-Fisher zero-point and $H_0$ vary depending on whether we use CPLR, TRGB or SNIa distances, and on which TRGB scale is adopted. In the $i$-band, the TRGB zero-point can vary by up to 0.14\,mag, corresponding to a variation in $H_0$ of $\pm$5\kmsMpc. If SNIa-derived distances anchored to Cepheid/TRGB calibrators are excluded from the absolute calibration, $H_0$ increases by 2.4\kmsMpc. \citet{Tully_2023} arrived at a similar finding when calibrating the absolute distance scale for the full CF4 catalogue. They found a lower $H_0$ value when calibrating to SNIa galaxies that hosted Cepheids, TRGB calibrators, or masers. This is in contrast to their preferred method of using an ensemble of secondary calibrators (including SNIa) with an arbitrary distance scale. Until there is a full consensus in the literature regarding the CPLR and TRGB calibrations, we will use the same calibrator set as K20 and adopt a systematic uncertainty on the Tully-Fisher zero-point of $\pm0.1$\,mag ($\pm$3.5\kmsMpc in $H_0$) to account for these conflicting CPLR/TRGB zero-points. There is ongoing discussion on this topic \citep{Freedman_2023, Madore_2023}, and we expect that the accuracy of this distance scale will continue to improve in the future.

There are also inconsistencies in the SNIa peak absolute magnitude $M_B$ between calibrations made with low-redshift data \citep{Riess_2022} and high-redshift data \citep{Dinda_2023, Camarena_2023, Camilleri_2024}. \citet{Camilleri_2024} used an `inverse distance ladder' technique, in which the SNIa distance scale is anchored to baryon acoustic oscillation (BAO) measurements at higher redshifts (primarily from the CMB), rather than low-redshift Cepheids. If we calibrate to SNIa distances only using the value of $M_B$ measured by \citet{Camilleri_2024}, this results in a zero-point that is brighter by 0.18\,mag (see Table~\ref{tab:zpvals}), and a correspondingly lower estimate of $H_0=66.5\pm2.1$\kmsMpc in the $i$-band. Unsurprisingly, this result is consistent with the $H_0$ estimated from the Planck CMB measurements \citep{2020_Planck}.

\begin{table*}
\centering
\caption{Tully-Fisher zero-point measurements, $a_0$, in magnitudes, and resulting $H_0$ values, in\,\kmsMpc, from various calibrator galaxy samples and absolute calibration methods/sources; $n$ is the number of calibrator galaxies in each sample. Unless otherwise indicated by a citation, the CPLR/TRGB zero-point scales are consistent with K20 and only galaxies with $\log\Wmxc>2$ are included. Note that the $a_0$ errors are only the statistical uncertainties from fitting a zero-point to the calibrator sample as in Section~\ref{sec:abscalib}, using the observational uncertainties of the absolute calibrators and assuming the other Tully-Fisher and peculiar velocity model parameters are exact. For the purposes of this calibration comparison, the $H_0$ uncertainties are just the statistical errors from a fit to the full CF4 sample as in Section~\ref{sec:H0measurement}, using the uncertainties given for the Tully-Fisher zero-point in the previous column.}
\label{tab:zpvals}
\begin{tabular}{lcccccc}
\hline\hline
\multirow{2}*{Absolute calibrators} &\multicolumn{3}{c}{SDSS $i$-band}&\multicolumn{3}{c}{WISE $W1$-band}\\
 &$a_0$&$H_0$&$n$&$a_0$&$H_0$&$n$\\
\hline
Cepheid & $-$21.00 $\pm$ 0.14 & 76.2 $\pm$ 4.9 & 16 & $-$20.54 $\pm$ 0.08 & 76.1 $\pm$ 3.0 & 22\\
TRGB & $-$21.03 $\pm$ 0.15 & 75.2 $\pm$ 5.0 & 23 & $-$20.54 $\pm$ 0.08 & 76.1 $\pm$ 2.7 & 38\\
SNIa & $-$21.12 $\pm$ 0.07 & 72.2 $\pm$ 2.3 & 61 & $-$20.61 $\pm$ 0.04 & 73.8 $\pm$ 1.4 & 75\\
CPLR+TRGB & $-$21.02 $\pm$ 0.11 & 75.7 $\pm$ 3.8 & 34 & $-$20.54 $\pm$ 0.06 & 76.3 $\pm$ 2.1 & 54\\
CPLR+TRGB+SNIa & $-$21.09 $\pm$ 0.06 & 73.3 $\pm$ 2.1 & 85 & $-$20.59 $\pm$ 0.03 & 74.5 $\pm$ 1.2 & 119\\
CPLR+TRGB+SNIa \& $\log\Wmxc$\,$>$\,2.2 & $-$21.09 $\pm$ 0.06 & 73.4 $\pm$ 2.1 & 85 & $-$20.58 $\pm$ 0.04 & 74.7 $\pm$ 1.2 & 119\\
TRGB \citep{Freedman_2019} & $-$21.17 $\pm$ 0.14 & 70.5 $\pm$ 4.7 & 23 & $-$20.68 $\pm$ 0.08 & 71.4 $\pm$ 2.6 & 38\\
SNIa \citep{Camilleri_2024} & $-$21.30 $\pm$ 0.07 & 66.5 $\pm$ 2.1 & 61 & $-$20.79 $\pm$ 0.04 & 68.0 $\pm$ 1.2 & 75\\
\hline
\end{tabular}
\end{table*}

Another choice that may affect the Tully-Fisher zero-point is the $\log\Wmxc$ cut. For the Tully-Fisher zero-point calibration in this analysis, we have included galaxies with $\log\Wmxc > 2$, which is roughly equivalent to the K20 selection of galaxies with absolute magnitude of $M<-17$ mag in the $i$ band and $M<-16.1$ mag in the $W1$ band. This is different to the lower limit adopted by \citet{Boubel_2024} of $\log\Wmxc > 2.2$, who found that this corresponds to the best trade-off between preserving data and minimising the scatter in the Tully-Fisher fit. However, since the calibrator sample is much smaller than the full CF4 sample, we prefer to widen the $\log\Wmxc$ range in order to include more galaxies and better constrain the zero-point. We find that the choice of $\log\Wmxc$ lower limit has almost no effect on the Tully-Fisher zero-point; Table~\ref{tab:zpvals} shows that the zero-points obtained using $\log\Wmxc > 2.2$ and $\log\Wmxc > 2$ are consistent with one another in both $i$-band and $W1$-band.

We also investigate the effect of statistical errors in the adopted peculiar velocity and Tully-Fisher model parameters on $H_0$ estimates. For each of the parameters $\parPV$ and $\parTF$ (excluding the zero-point $a_0$, since that is being determined separately), we draw from Gaussian distributions with 1$\sigma$ uncertainties and mean values taken from Table~\ref{tab:params} and generate 1000 different $H_0$ values in each band. This exercise assumes symmetric, Gaussian, and uncorrelated distributions for each parameter. For the $i$-band the RMS variation of $H_0$ is 0.20\kmsMpc\ and for the $W1$-band the RMS in $H_0$ is 0.07\kmsMpc. For the calibration samples, the statistical uncertainties in the Tully-Fisher zero-points $a_0$ are larger ($\pm$0.06 mag in the $i$-band and $\pm$0.03 mag in the $W1$ band; from Table~\ref{tab:zpvals}) due to the smaller number of calibrator galaxies. We estimate the corresponding statistical uncertainties in $H_0$ to be $\pm$2.0\kmsMpc\ for the $i$-band and $\pm$1.2\kmsMpc\ for the $W1$-band. These are the dominant statistical uncertainties, and can only be reduced by increasing the number of absolute distance calibrators.

Another potential source of error for the Tully-Fisher relation lies in measurements of galaxy inclinations, as velocity widths are computed from $\Wmxc = W_\textrm{mx}/\sin{i}$, where $W_\textrm{mx}$ is the measured line width and $i$ is the measured inclination. Large errors in $i$ would result in asymmetric uncertainties on $\log\Wmxc$, which can potentially lead to a biased result (and in fact the errors in $i$ may themselves be asymmetric). \citet{Kourkchi_2020} measured galaxy inclinations using an online citizen-science tool they developed. They determined that the uncertainties were 2--5$\degree$, depending on the inclination. In our analysis thus far, we have propagated these inclination uncertainties into the uncertainties of $\log\Wmxc$, increasing the errors on $\log\Wmxc$ by up to 0.03 dex (published CF4 uncertainties in $\log\Wmxc$ appear to include only errors in $\log\Wmx$ and not errors in $i$). We note that, for both bands, this results in a final estimate of $H_0$ that is lower by 0.17\kmsMpc and has an additional statistical error of about 0.1\kmsMpc compared to one in which inclination errors are ignored. To confirm that this statistical uncertainty and potential systematic bias on $H_0$ arise directly from inclination errors, we repeated the calibration of the Tully-Fisher parameters and calculation of $H_0$, but with 1500 different sets of $\log\Wmxc$. We shift $\log\Wmxc$ by amounts corresponding to variations of $i$ drawn randomly from a Gaussian distribution with RMS given by the uncertainty in the inclination of each galaxy. We find that the RMS variation of $H_0$ is 0.11\kmsMpc\ for the $i$ band and 0.15\kmsMpc\ for the $W1$ band. These results are consistent with simply propagating the inclination errors and demonstrates that even a symmetric uncertainty in inclination will result in a small bias in $H_0$.

\section{Conclusions}
\label{sec:conclusions}

We present an improved method for measuring $H_0$ using the Tully-Fisher relation and a peculiar velocity model. y-Fisher relation and peculiar velocity model are fit simultaneously using the full sample of galaxies with suitable line-widths and magnitudes following the method of \citet{Boubel_2024}. This enables us to obtain more precise estimates for the cosmological redshift of each galaxy. The Tully-Fisher relation zero-point calibration is performed separately using a subset of galaxies that have distances derived from absolute distance estimators, allowing the estimation of inferred distances for the entire Tully-Fisher sample. These redshifts and distances, and their associated errors, are then used to estimate $H_0$.

The Cosmicflows-4 $i$-band Tully-Fisher sample yields a Hubble constant $H_0$\,=\,73.3\,$\pm$\,2.1\,(stat)\,$\pm$\,3.5\,(sys)\kmsMpc\ based on an \textit{absolute} zero-point calibration rom the subset of galaxies with distances obtained from the Cepheid period--luminosity relation, the tip of the red giant branch, and/or the supernovae Type Ia standard candle indicators. A comparison to the calibrated Tully-Fisher relation of K20 produces a \textit{relative} zero-point calibration for the $i$-band Tully-Fisher relation that yields  $H_0$\,=\,74.3\,$\pm$\,2.2\,(stat)\,$\pm$\,2.3\,(sys)\kmsMpc. For the Cosmicflows-4 $W1$-band sample, an absolute zero-point calibration gives $H_0$\,=\,74.5\,$\pm$\,1.2\,(stat)\,$\pm$\,2.6\,(sys)\kmsMpc\ and a relative calibration gives $H_0$\,=\,73.2\,$\pm$\,4.8\,(stat)\,$\pm$\,2.3\,(sys)\kmsMpc. 

\begin{figure}
\centering
\includegraphics[width=1.0\columnwidth]{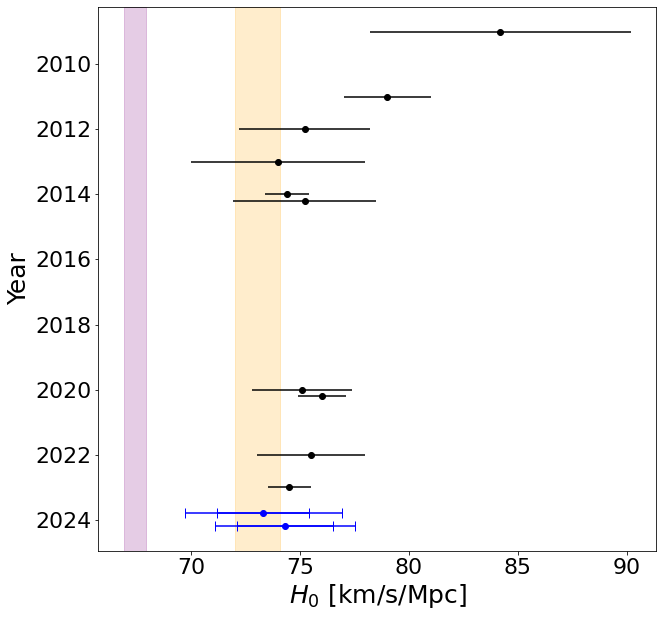}
\caption{Published estimates of $H_0$ from the Tully-Fisher relation, starting from 2009. For comparison, the purple band shows the \citet{2020_Planck} value, $H_0$\,=\,67.4\,$\pm$\,0.5\kmsMpc, and the orange band the \citet{Riess_2022} value, $H_0$\,=\,73.0\,$\pm$\,1.0\kmsMpc. References and $H_0$ values are listed in Table~\ref{tab:H0tf}. The blue points are from this work, using the absolute calibration (upper point) and the relative calibration (lower point) in the $i$-band; the shorter error bars correspond to statistical uncertainties only.}
\label{fig:h0_compare}
\end{figure}

Figure~\ref{fig:h0_compare} compares our measured values of $H_0$ to those of other studies over the last 15 years that also used the Tully-Fisher relation. Much of the data is common between these studies, but they differ in the methods used. The measurements presented here are unique in that they: incorporate the largest amount of data; employ a comprehensive Tully-Fisher model with curvature and varying intrinsic scatter; exploit a peculiar velocity model based on redshift surveys; account for selection effects and Malmquist bias; and are relatively simple, inasmuch as the method does not necessitate additional corrections and involves only a small number of steps. However, the derived overall uncertainties on $H_0$ are large compared to some of the other measurements. This is due in part to the comprehensive inclusion of all sources of error, but primarily to the systematic uncertainty attributed to the choice of absolute distance calibrators (CPLR versus TGRB), given the controversy in the literature regarding their corresponding zero-point scales \citep{Freedman_2019, Yuan_2019, Riess_2019, Freedman_2020, Riess_2022, Freedman_2023}. The systematic uncertainty in the absolute distance calibrators is presently the limiting factor in obtaining a higher-precision measurement, followed by the statistical uncertainty from the relatively small size of the sample with absolute distance estimates. As evidenced by numerous papers over the last few years, considerable research effort is going into these issues and some progress is being made. With consensus on the absolute distance calibrations, and a larger number of galaxies having such distances, the advantages of our method for estimating $H_0$ from the Tully-Fisher relation will become significant.

In addition to the unreliability of absolute distance indicators, other limitations in using the Tully-Fisher relation include the difficulty in accurately measuring galaxy inclinations and the large intrinsic scatter of the Tully-Fisher relation. A recent study \citep{Fu_2024} presented a method for correcting Tully-Fisher data for inclinations without the need to measure individual inclinations and their errors. While the methodology advocated here does not present solutions to those issues, it provides a far more robust Tully-Fisher template than previous methods. With the expected bounty of Tully-Fisher data from new surveys such as WALLABY \citep{Koribalski_2020, Westmeier_2022}, our method will enhance the use of the Tully-Fisher relation as a key tool for precise measurements of $H_0$ at low redshifts.

\section*{Acknowledgements}

MMC acknowledges support from a Royal Society Wolfson Visiting Fellowship (RSWVF{\textbackslash}R3{\textbackslash}223005) at the University of Oxford. KS acknowledges support through the Australian Research Council’s Laureate Fellowship funding scheme (project FL180100168). We acknowledge use of the following analysis packages: Astropy \citep{astropy}, GetDist \citep{getdist}, emcee \citep{emcee}, and Matplotlib \citep{matplotlib}.

\section*{Data Availability}

This work uses previously published data as referenced and described in text. Details of the analysis and code can be made available upon reasonable request to the corresponding author.



\bibliographystyle{mnras}
\bibliography{sources}

\begin{thebibliography}{}
\makeatletter
\relax
\def\mn@urlcharsother{\let\do\@makeother \do\$\do\&\do\#\do\^\do\_\do\%\do\~}
\def\mn@doi{\begingroup\mn@urlcharsother \@ifnextchar [ {\mn@doi@} {\mn@doi@[]}}
\def\mn@doi@[#1]#2{\def\@tempa{#1}\ifx\@tempa\@empty \href {http://dx.doi.org/#2} {doi:#2}\else \href {http://dx.doi.org/#2} {#1}\fi \endgroup}
\def\mn@eprint#1#2{\mn@eprint@#1:#2::\@nil}
\def\mn@eprint@arXiv#1{\href {http://arxiv.org/abs/#1} {{\tt arXiv:#1}}}
\def\mn@eprint@dblp#1{\href {http://dblp.uni-trier.de/rec/bibtex/#1.xml} {dblp:#1}}
\def\mn@eprint@#1:#2:#3:#4\@nil{\def\@tempa {#1}\def\@tempb {#2}\def\@tempc {#3}\ifx \@tempc \@empty \let \@tempc \@tempb \let \@tempb \@tempa \fi \ifx \@tempb \@empty \def\@tempb {arXiv}\fi \@ifundefined {mn@eprint@\@tempb}{\@tempb:\@tempc}{\expandafter \expandafter \csname mn@eprint@\@tempb\endcsname \expandafter{\@tempc}}}

\bibitem[\protect\citeauthoryear{Boubel, Colless, Said  \& Staveley-Smith}{Boubel et~al.}{2024}]{Boubel_2024}
Boubel P.,  Colless M.,  Said K.,   Staveley-Smith L.,  2024, MNRAS, 531, 84

\bibitem[\protect\citeauthoryear{Camarena \& Marra}{Camarena \& Marra}{2023}]{Camarena_2023}
Camarena D.,  Marra V.,  2023, The tension in the absolute magnitude of Type Ia supernovae (\mn@eprint {arXiv} {2307.02434})

\bibitem[\protect\citeauthoryear{Camilleri et~al.,}{Camilleri et~al.}{2024}]{Camilleri_2024}
Camilleri R.,  et~al., 2024, The Dark Energy Survey Supernova Program: An updated measurement of the Hubble constant using the Inverse Distance Ladder (\mn@eprint {arXiv} {2406.05049})

\bibitem[\protect\citeauthoryear{{Chiba} \& {Nakamura}}{{Chiba} \& {Nakamura}}{1998}]{Chiba_1998}
{Chiba} T.,  {Nakamura} T.,  1998, Progress of Theoretical Physics, 100, 1077

\bibitem[\protect\citeauthoryear{Courtois, Dupuy, Guinet, Baulieu, Ruppin  \& Brenas}{Courtois et~al.}{2023}]{Courtois_2023}
Courtois H.~M.,  Dupuy A.,  Guinet D.,  Baulieu G.,  Ruppin F.,   Brenas P.,  2023, A\&A, 670, L15

\bibitem[\protect\citeauthoryear{Dinda \& Banerjee}{Dinda \& Banerjee}{2023}]{Dinda_2023}
Dinda B.~R.,  Banerjee N.,  2023, Phys. Rev. D, 107, 063513

\bibitem[\protect\citeauthoryear{Foreman-Mackey, Hogg, Lang  \& Goodman}{Foreman-Mackey et~al.}{2013}]{emcee}
Foreman-Mackey D.,  Hogg D.,  Lang D.,   Goodman J.,  2013, PASP, 125, 306

\bibitem[\protect\citeauthoryear{Freedman \& Madore}{Freedman \& Madore}{2023}]{Freedman_2023}
Freedman W.~L.,  Madore B.~F.,  2023, JCAP, 2023, 050

\bibitem[\protect\citeauthoryear{{Freedman} et~al.,}{{Freedman} et~al.}{2019}]{Freedman_2019}
{Freedman} W.~L.,  et~al., 2019, \apj, 882, 34

\bibitem[\protect\citeauthoryear{Freedman et~al.,}{Freedman et~al.}{2020}]{Freedman_2020}
Freedman W.~L.,  et~al., 2020, ApJ, 891, 57

\bibitem[\protect\citeauthoryear{Fu}{Fu}{2024}]{Fu_2024}
Fu H.,  2024, The Astrophysical Journal Letters, 963, L19

\bibitem[\protect\citeauthoryear{Hislop et~al.,}{Hislop et~al.}{2011}]{Hislop_2011}
Hislop L.,  et~al., 2011, ApJ, 733, 75

\bibitem[\protect\citeauthoryear{Hoyt}{Hoyt}{2021}]{Hoyt_2021}
Hoyt T.~J.,  2021, On Zero Point Calibration of the Red Giant Branch Tip in the Magellanic Clouds (\mn@eprint {arXiv} {2106.13337})

\bibitem[\protect\citeauthoryear{{Hunter}}{{Hunter}}{2007}]{matplotlib}
{Hunter} J.~D.,  2007, Comput. Sci. Eng., 9, 90

\bibitem[\protect\citeauthoryear{Koribalski et~al.,}{Koribalski et~al.}{2020}]{Koribalski_2020}
Koribalski B.~S.,  et~al., 2020, ApSS, 365

\bibitem[\protect\citeauthoryear{Kourkchi, Tully, Anand, Courtois, Dupuy, Neill, Rizzi  \& Seibert}{Kourkchi et~al.}{2020}]{Kourkchi_2020}
Kourkchi E.,  Tully R.~B.,  Anand G.~S.,  Courtois H.~M.,  Dupuy A.,  Neill J.~D.,  Rizzi L.,   Seibert M.,  2020, ApJ, 896, 3

\bibitem[\protect\citeauthoryear{Kourkchi, Tully, Courtois, Dupuy  \& Guinet}{Kourkchi et~al.}{2022}]{Kourkchi_2022}
Kourkchi E.,  Tully R.~B.,  Courtois H.~M.,  Dupuy A.,   Guinet D.,  2022, MNRAS, 511, 6160

\bibitem[\protect\citeauthoryear{Lewis}{Lewis}{2019}]{getdist}
Lewis A.,  2019, GetDist: a Python package for analysing Monte Carlo samples

\bibitem[\protect\citeauthoryear{Madore, Freedman  \& Owens}{Madore et~al.}{2023}]{Madore_2023}
Madore B.~F.,  Freedman W.~L.,   Owens K.,  2023, AJ, 166, 224

\bibitem[\protect\citeauthoryear{Makarov, Prugniel, Terekhova, Courtois  \& Vauglin}{Makarov et~al.}{2014}]{Makarov_2014}
Makarov D.,  Prugniel P.,  Terekhova N.,  Courtois H.,   Vauglin I.,  2014, A\&A, 570, A13

\bibitem[\protect\citeauthoryear{Neill, Seibert, Tully, Courtois, Sorce, Jarrett, Scowcroft  \& Masci}{Neill et~al.}{2014}]{Neill_2014}
Neill J.~D.,  Seibert M.,  Tully R.~B.,  Courtois H.,  Sorce J.~G.,  Jarrett T.~H.,  Scowcroft V.,   Masci F.~J.,  2014, ApJ, 792, 129

\bibitem[\protect\citeauthoryear{Planck~Collaboration: Aghanim et~al.,}{Planck~Collaboration: et~al.}{2020}]{2020_Planck}
Planck~Collaboration: Aghanim N.,  et~al., 2020, \aap, 641, A6

\bibitem[\protect\citeauthoryear{Riess, Casertano, Yuan, Macri  \& Scolnic}{Riess et~al.}{2019}]{Riess_2019}
Riess A.,  Casertano S.,  Yuan W.,  Macri L.,   Scolnic D.,  2019, ApJ, 876, 85

\bibitem[\protect\citeauthoryear{Riess et~al.,}{Riess et~al.}{2022}]{Riess_2022}
Riess A.~G.,  et~al., 2022, ApJL, 934, L7

\bibitem[\protect\citeauthoryear{{Rizzi}, {Tully}, {Makarov}, {Makarova}, {Dolphin}, {Sakai}  \& {Shaya}}{{Rizzi} et~al.}{2007}]{Rizzi_2007}
{Rizzi} L.,  {Tully} R.~B.,  {Makarov} D.,  {Makarova} L.,  {Dolphin} A.~E.,  {Sakai} S.,   {Shaya} E.~J.,  2007, \apj, 661, 815

\bibitem[\protect\citeauthoryear{Robitaille et~al.,}{Robitaille et~al.}{2013}]{astropy}
Robitaille T.~P.,  et~al., 2013, \aap, 558, A33

\bibitem[\protect\citeauthoryear{Russell}{Russell}{2009}]{Russell_2009}
Russell D.~G.,  2009, Journal of Astrophysics and Astronomy, 30, 93–118

\bibitem[\protect\citeauthoryear{Schombert, McGaugh  \& Lelli}{Schombert et~al.}{2020}]{Schombert_2020}
Schombert J.,  McGaugh S.,   Lelli F.,  2020, AJ, 160, 71

\bibitem[\protect\citeauthoryear{Sorce, Tully  \& Courtois}{Sorce et~al.}{2012}]{Sorce_2012}
Sorce J.~G.,  Tully R.~B.,   Courtois H.~M.,  2012, ApJL, 758, L12

\bibitem[\protect\citeauthoryear{Sorce et~al.,}{Sorce et~al.}{2013}]{Sorce_2013}
Sorce J.~G.,  et~al., 2013, ApJ, 765, 94

\bibitem[\protect\citeauthoryear{{Sorce}, {Tully}, {Courtois}, {Jarrett}, {Neill}  \& {Shaya}}{{Sorce} et~al.}{2014}]{Sorce_2014}
{Sorce} J.~G.,  {Tully} R.~B.,  {Courtois} H.~M.,  {Jarrett} T.~H.,  {Neill} J.~D.,   {Shaya} E.~J.,  2014, MNRAS, 444, 527

\bibitem[\protect\citeauthoryear{Strauss \& Willick}{Strauss \& Willick}{1995}]{Strauss_1995}
Strauss M.~A.,  Willick J.~A.,  1995, Phys. Rep., 261, 271

\bibitem[\protect\citeauthoryear{{Tully} \& {Fisher}}{{Tully} \& {Fisher}}{1977}]{Tully_1977}
{Tully} R.~B.,  {Fisher} J.~R.,  1977, \aap, 54, 661

\bibitem[\protect\citeauthoryear{Tully et~al.,}{Tully et~al.}{2023}]{Tully_2023}
Tully R.~B.,  et~al., 2023, The Astrophysical Journal, 944, 94

\bibitem[\protect\citeauthoryear{{Visser}}{{Visser}}{2004}]{Visser_2004}
{Visser} M.,  2004, Classical and Quantum Gravity, 21, 2603

\bibitem[\protect\citeauthoryear{Westmeier et~al.,}{Westmeier et~al.}{2022}]{Westmeier_2022}
Westmeier T.,  et~al., 2022, PASA, 39

\bibitem[\protect\citeauthoryear{{Yuan}, {Riess}, {Macri}, {Casertano}  \& {Scolnic}}{{Yuan} et~al.}{2019}]{Yuan_2019}
{Yuan} W.,  {Riess} A.~G.,  {Macri} L.~M.,  {Casertano} S.,   {Scolnic} D.~M.,  2019, \apj, 886, 61

\makeatother
\end{thebibliography}



\bsp	
\label{lastpage}
\end{document}